\documentstyle[prb,aps,twocolumn,graphicx]{revtex}

\begin{document}
\draft
\twocolumn [\hsize\textwidth\columnwidth\hsize\csname
@twocolumnfalse\endcsname

\title{Evolution of spin excitations in a gapped antiferromagnet from the
quantum to the high-temperature limit}

\author{M. Kenzelmann,$^{1}$ R.~A. Cowley,$^{1}$ W.~J.~L.
Buyers,$^{2,3}$ R. Coldea,$^{4,5}$ M. Enderle$^{6,7}$ and D.~F.
McMorrow$^{8}$}

\address{(1) Oxford Physics, Clarendon Laboratory, Oxford OX1 3PU, United Kingdom \\(2)
Neutron Program for Materials Research, National Research Council
of Canada, Chalk River, Ontario, Canada KOJ 1J0 \\(3) Canadian
Institute for Advanced Research \\(4) Oak Ridge National
Laboratory, Solid State Division, Oak Ridge, Tennessee 37831\\(5)
ISIS Facility, Rutherford Appleton Laboratory, Oxon OX11 0QX,
United Kingdom\\(6) Technische Physik, Geb\"{a}ude 38,
Universit\"{a}t des Saarlandes, 66123 Saarbr\"{u}cken,
Germany\\(7) Institut Laue-Langevin, BP 156 38042 Grenoble, Cedex
9, France\\(8) Condensed Matter Physics and Chemistry Department,
$Ris\o$ National Laboratory, DK-4000, Roskilde, Denmark}
\date{\today}
\maketitle

\begin{abstract}
We have mapped from the quantum to the classical limit the spin
excitation spectrum of the antiferromagnetic spin-1 Heisenberg
chain system ${\rm CsNiCl_{3}}$ in its paramagnetic phase from
$T=5$ to $200\;\mathrm{K}$. Neutron scattering shows that the
excitations are resonant and dispersive up to at least
$T=70\;\mathrm{K}$, but broaden considerably with increasing
temperature. The dispersion flattens out with increasing
temperature as the resonance energy $\Delta$ at the
antiferromagnetic wave-vector increases and the maximum in the
dispersion decreases. The correlation length $\xi$ between $T=12$
and $50\;\mathrm{K}$ is in agreement with quantum Monte Carlo
calculations. $\xi$ is also consistent with the single mode
approximation, suggesting that the excitations are short-lived
single particle excitations. Below $T=12\;\mathrm{K}$ where
three-dimensional spin correlations are important, $\xi$ is
shorter than predicted and the experiment is not consistent with
the random phase approximation for coupled quantum chains. At
$T=200\;\mathrm{K}$, the structure factor and second energy moment
of the excitation spectrum are in excellent agreement with the
high-temperature series expansion.
\end{abstract}
 ]

\newpage

\section{Introduction}
Low-dimensional antiferromagnets are ideal systems in which to
investigate the effect of strong quantum fluctuations on the
dynamics of interacting spin systems. Quantum fluctuations are
similar to temperature fluctuations in three-dimensional (3D)
magnets in that they disorder the spin system and prevent magnetic
order. Like the paramagnetic phase of a 3D magnet, the
quantum-disordered phase of a low-dimensional antiferromagnet has
unbroken translational symmetry, but while the excitation spectrum
of a 3D paramagnet consists of broad magnetic excitations, the
elementary magnetic excitations of a low-dimensional
antiferromagnet are spin-1/2 or long-lived spin-1
particles.\cite{Faddeev_Takhatajan,Haldane83,Haldane93}

One-dimensional (1D) antiferromagnets have been studied for at
least the last 30 years. They were first perceived as simpler
model systems than their 3D counterparts, but it is now clear that
their low-temperature spin dynamics is far more complex. The
quantum-disordered ground state of 1D antiferromagnets is usually
a highly correlated spin singlet. Haldane, in his seminal 1983
paper, conjectured that the spin dynamics of 1D antiferromagnets
crucially depends on the spin quantum number:\cite{Haldane83} The
half-integer spin Heisenberg antiferromagnet has a gapless
excitation spectrum while integer-spin chains exhibit an energy
gap. Haldane's conjecture of a spin-1 gap was then  confirmed by
numerical\cite{Botet_Julien,Nightingale_Blote} and experimental
studies\cite{Buyers86,Steiner,Renard} and is now generally
accepted.\par

The low-temperature quantum-disordered ground state of the
antiferromagnetic spin-1/2 chain is "almost ordered" and its
correlations fall off with distance as a power law. The elementary
excitations are spinons carrying spin-1/2 which correspond to
moving domain walls between regions of coherent quantum states. In
contrast, the correlations in antiferromagnetic spin-1 chains fall
off exponentially with a correlation length of about 6 sites at
zero temperature. The elementary excitations are massive
triply-degenerate spin-1 particles, called Haldane excitations,
which can be pictured as domain walls breaking the hidden spin
string order that is inherent in spin-1 chains.\cite{Fath_Solyom}
The triplet excitations follow the relativistic dispersion of the
non-linear sigma model (${\rm NL\sigma M}$).\cite{Affleck}\par

The temperature dependence of the excitation spectrum of
antiferromagnetic chains has been measured mostly for $T \leq J$,
where $J$ is the exchange interaction between the spins along the
chain direction. For ${\rm KCuF_3}$, an antiferromagnetic spin-1/2
chain system with $J=17\;\mathrm{meV}$, inelastic neutron
scattering up to $T=200\;{\mathrm{K}}\simeq J$ showed that the
low-energy excitations are broadened at high temperatures but can
still be described by a low-temperature field
theory.\cite{Tennant93}\par

For spin-1 chains, the temperature dependence of
$S(\mathbf{Q},\omega)$ has been investigated for ${\rm
CsNiCl_3}$\cite{Buyers86,William_Buyers_4,Steiner,Zaliznyak},
NENP\cite{Renard}, ${\rm AgVP_2S_6}$\cite{Mutka_AgVP2S6} and ${\rm
Y_2BaNiO_5}$\cite{Darriet} for temperatures $T$ up to at most $T
\sim J$. The temperature dependence of the excitation for $T<0.3J$
was studied for ${\rm AgVP_2S_6}$ and ${\rm Y_2BaNiO_5}$
($J=60\;\mathrm{meV}$ and $J=24.1\;\mathrm{meV}$, respectively).
These measurements give the behavior of the low-temperature
quantum-disordered phase. A further inelastic neutron scattering
study of ${\rm AgVP_2S_6}$ showed that the staggered
susceptibility decreases with increasing temperature up to
$T=200\;{\mathrm{K}} \simeq 0.3 J$.\cite{Mutka_comparison} A
single crystal experiment on ${\rm Y_2BaNiO_5}$ showed that the
energy gap observed by neutron scattering is renormalized upward
with increasing temperature up to $T=80\;{\mathrm{K}} \simeq 0.3
J$.\cite{Sakaguchi}\par

${\rm CsNiCl_3}$ and NENP have much smaller exchange interactions,
$J=2.28\;\mathrm{meV}$ and $J=4\;\mathrm{meV}$, respectively, and
were studied over a wider temperature range. In NENP the upward
renormalization of the energy gap in the spectrum was measured up
to $T=20\;{\mathrm{K}} \simeq 0.43 J$,\cite{Renard_NENP_NINO} and
the temperature dependence of the correlation length has been
determined up to $T=100\;{\mathrm{K}} \simeq 2 J$.\cite{Ma95} The
temperature dependence of the gap in ${\rm CsNiCl_3}$ at the 3D
ordering wave-vector was previously studied up to
$T=20\;{\mathrm{K}}\simeq 0.75 J$,\cite{Steiner,Zaliznyak} and at
the 1D wave-vector up to $T=12\;{\mathrm{K}}\simeq 0.5
J$.\cite{Morra,William_Buyers_4}\par

Surprisingly a detailed study of the excitation spectrum of
antiferromagnetic spin chains for $T > J$ and in the
high-temperature limit has not been made. We showed recently that
the Haldane gap in ${\rm CsNiCl_3}$ persists as a resonant feature
up to at least $T=70\;\mathrm{K}$ or
$2.7J$.\cite{Kenzelmann_CsNiCl3_gap} Here we present more
extensive measurements at low and high temperatures including a
comprehensive study of the spin dynamics of the 1D magnet for
temperatures $T > J$. ${\rm CsNiCl_{3}}$ has an exchange
interaction $J$ that makes it an ideal system for studying
high-temperature fluctuations, not only because the whole magnetic
excitation spectrum can be measured with thermal neutrons but also
because temperatures between $J$ and $\sim 10 J$ (room
temperature) can readily be accessed.\par

Section II of the paper describes magnetic properties of ${\rm
CsNiCl_3}$ and gives the configuration of the neutron scattering
spectrometers. Section III gives a thorough analysis of the
measurements presented earlier.\cite{Kenzelmann_CsNiCl3_gap} The
inelastic time-of-flight neutron scattering measurements of the
excitation spectrum between $T=12$ and $50\;\mathrm{K}$ are
reported in Section IV. In this temperature range, resonant and
dispersive excitations were observed for a wide range of
wave-vector transfers along the chain. In Section V, we describe
the measurement of the dynamic structure factor,
$S(\bbox{Q},\omega)$, of the paramagnetic scattering at
$T=200\;\mathrm{K}$. The results are summarized, discussed and put
into an overall context in Section VI.

\section{Experimental details}
${\rm CsNiCl_{3}}$ is one of the best examples of a gapped
weakly-coupled spin-1 chain system. The crystal structure is
hexagonal, $D^{4}_{6h}$ space group, with low-temperature lattice
constants $a=7.14\,$\AA \, and $c = 5.90\,$\AA \,. The exchange
interaction along the c-axis, $J=2.28\;\mathrm{meV}$,\cite{Katori}
is much stronger than the interaction in the basal plane,
$J'=0.044\;\mathrm{meV}$,\cite{Buyers86,Morra} making ${\rm
CsNiCl_{3}}$ a system of weakly-coupled spin-1 chains. The
Hamiltonian can be written as
\begin{equation}
{\mathcal{H}}=J \sum_{i}^{\rm chain} \bbox{S}_{\rm i} \cdot
\bbox{S}_{\rm i+1}+ J' \sum_{<i,j>}^{\rm plane} \bbox{S}_{\rm i}
\cdot \bbox{S}_{\rm j} - D \sum_{i} (S_{\rm i}^{z})^2\, .
\label{Hamiltonian}
\end{equation}
The weak Ising anisotropy $D=4\;\mathrm{\mu eV}$ is small enough
that ${\rm CsNiCl_{3}}$ is a good example of an isotropic
Heisenberg antiferromagnet.\par

Because of the interchain couplings ${\rm CsNiCl_{3}}$ orders
antiferromagnetically below $T_{\rm N}=
4.84\;\mathrm{K}$.\cite{Morra} Above $T_{\rm N}$, ${\rm
CsNiCl_{3}}$ displays the characteristics of an isolated
antiferromagnetic spin-1 chain: the excitation spectrum exhibits
the Haldane gap, the correlations fall off exponentially and the
magnetic susceptibility shows a broad maximum near
$30\;\mathrm{K}$.\cite{Achiwa}\par

The single crystal of ${\rm CsNiCl_{3}}$ $5\;\mathrm{mm} \times
5\;\mathrm{mm} \times 20\;\mathrm{mm}$ was sealed in an aluminum
can containing helium gas to prevent absorption of water. The
experiments were performed using the cold-neutron RITA triple axis
spectrometer\cite{Lefman} at the DR3 reactor of the ${\rm Ris\o}$
National Laboratory, Denmark, and the chopper time-of-flight
spectrometer MARI at the pulsed spallation source ISIS of the
Rutherford Appleton Laboratory, United Kingdom.\par

For the experiment using the RITA spectrometer, the sample was
mounted in a cryostat with its (hhl) crystallographic plane in the
horizontal scattering plane of the neutron spectrometer, and the
temperature was controlled to an accuracy of $0.1\;\mathrm{K}$
between $1.5$ and $70\;\mathrm{K}$. The energy of the neutrons
from a cold source was selected by a vertically focusing pyrolytic
graphite monochromator. A rotating velocity selector before the
monochromator suppressed unwanted neutrons that would be Bragg
reflected by its higher order planes. Supermirror guides with
$n=3$ ($\theta_{\rm c}=1.2^{\rm o}$ at $4$\,\AA) were located
before and after the monochromator, and the beam was
$20\;\mathrm{mm}$ wide. The scattered beam was filtered through
cooled beryllium, passed through a $50\;\mathrm{mm}$ wide $1^{\rm
o}$ Soller collimator, and was analyzed by reflection from the
central two blades of a 7-component flat pyrolytic graphite
analyzer aligned so that each blade reflected the same energy of
neutrons, $5\;\mathrm{meV}$. The Soller geometry meant that the
beam reflected by the analyzers was about $20\;\mathrm{mm}$ wide
and located in the central 30 strips of a position sensitive
detector. Because the detector was 120 pixels wide
($\sim120\;\mathrm{mm}$) the side strips were used to estimate and
subtract the temperature independent background. Turning of either
analyzer or monochromator from their reflecting position confirmed
that the side pixels gave a good representation of the background
arriving at the centre of the detector. With this arrangement the
low-energy resolution was typically $0.26\;\mathrm{meV}$ (FWHM) in
energy. The longitudinal and transverse wave-vector width of the
$(002)$-peak was $0.01$ and $0.016$ (FWHM in reciprocal lattice
units) and the calculated vertical resolution was
$0.22$\,\AA$^{-1}$.\par

For the experiment using the MARI spectrometer the sample was
oriented with its (hhl) plane in the vertical scattering plane,
and the c-axis was either parallel or perpendicular to the
incident beam direction. The incident neutron energy was either
$20$ or $30\;\mathrm{meV}$ with the Fermi chopper spinning at
$150\;\mathrm{Hz}$. The energy resolution was then $0.35$ or
$0.4\;\mathrm{meV}$, respectively, as determined from the full
width at half maximum (FWHM) of the quasi-elastic peak. The
resolution in wave-vector transfer at zero energy transfer was
typically $0.02$\,\AA$^{-1}$ along both the $c^{\star}$ and the
$[110]$ directions and as large as $0.19$\,\AA$^{-1}$
perpendicular to the scattering plane if only the central detector
bank was used. Both the energy and the wave-vector resolution
improved with increasing energy transfer. The measurements were
performed with the sample at temperatures between $T=6.2$ and
$200\;\mathrm{K}$. The scattering was measured for a total proton
charge between $4000$ and $8600\;\mathrm{\mu Ah}$ at an average
proton current of $~170\;\mathrm{\mu A}$.\par

MARI has three detector banks two of which are out of the vertical
scattering plane. For $T\leq25\;\mathrm{K}$ only data recorded by
the central detector bank was included in the analysis because the
two side banks sample an out-of-plane wave-vector transfer where
the neutron scattering is different. All three detector banks were
used for the measurements performed at temperatures $T \geq
50\;\mathrm{K}$ where the 3D dispersion is small and good
resolution perpendicular to the scattering plane is
unnecessary.\par

\section{Experimental results}

\subsection{Excitation at ${\bbox Q_c=1}$}\label{Section-pi}
Close to the antiferromagnetic point, denoted $Q_c=1$ because the
spin spacing is $c/2$ along the chain, the excitation spectrum of
an antiferromagnetic spin-1 chain in its quantum-disordered phase
is dominated by well-defined, gapped spin-1 excitations (Haldane
excitations) which satisfy a relativistic dispersion relation. We
showed in a recent publication\cite{Kenzelmann_CsNiCl3_gap} that
for $T \leq \Delta$, the Haldane excitation at $Q_c=1$ exhibited a
Lorentzian spectral form. We showed that its life-time agrees well
with a semi-classical theory\cite{Damle_Sachdev} for gapped
systems provided we replaced the classical dispersion by the more
accurate relativistic dispersion of the ${\rm NL\sigma M}$. At
higher temperatures the Haldane excitation evolves into a broad
resonant feature. In this paper a more detailed analysis of the
data is presented and results based on two different spectral
functions are presented for temperatures between $T=5$ and
$T=70\;\mathrm{K}=2.7J$.

\begin{figure}
\begin{center}
  \includegraphics[height=6cm,bbllx=50,bblly=335,bburx=472,
  bbury=640,angle=0,clip=]{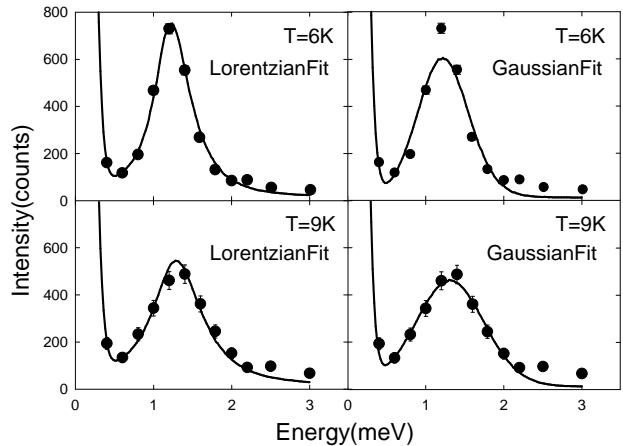}
  \vspace{0.3cm}
  \caption{Scattering from the Haldane excitation at
  $\bbox{Q}=(0.81,0.81,1)$ for $T=6$ and $9\;\mathrm{K}$ as
  a function of energy transfer. The data were measured using
  the triple-axis spectrometer RITA as described in the text.
  The intensity was fitted to a Lorentzian (left) and a Gaussian
  (right) cross-section convoluted with the resolution
  function.\protect\cite{Cooper_Nathans}}
  \label{lineshape}
\end{center}
\end{figure}

The Haldane excitation was measured at $\bbox{Q}=(0.81,0.81,1)$
where the Fourier transform of the interchain coupling vanishes,
and within a random phase approximation (RPA) the chains behave as
if they were uncoupled.\cite{Buyers86} In Fig.~\ref{lineshape},
the results are shown at $T=6$ and $9\;\mathrm{K}$. The spectra
were fitted to both an antisymmetrized Lorentzian and an
antisymmetrized Gaussian line-shape, weighted by the Bose factor,
and convoluted with the resolution function.\cite{Cooper_Nathans}
It is clear from the spectrum at $T=6\;\mathrm{K}$ that an
intrinsic Lorentzian line-shape gives a better description of the
measured data, as predicted by Damle and
Sachdev.\cite{Damle_Sachdev}\par

The Lorentzian fits were performed with a double Lorentzian form,
namely:
\begin{eqnarray}
        S(\bbox{Q},\omega)= A \cdot \left( n(\omega)+1 \right)
        \cdot \hspace{1cm} \nonumber \\ \left(\frac{\Gamma}
        {(\omega-\epsilon(\bbox{Q}))^2 +\Gamma^2} - \frac{\Gamma}
        {(\omega+\epsilon(\bbox{Q}))^2+\Gamma^2}\right)\, ,
        \label{Lorentzian_form}
\end{eqnarray}where $\omega$ is the energy transfer and the Bose
factor is defined by
\begin{equation}
    n(\omega)+1=(1-\exp(-\frac{\hbar\omega}{k_B T}))^{-1}\, .
\end{equation}The energy width is given by $\Gamma$
(half Lorentzian width at half maximum) and the energy of the
excitations by $\epsilon(\bbox{Q})$. $A$ is an overall scaling
factor which scales as $1/\epsilon(\bbox{Q})$ for
antiferromagnetic spin waves and also for the excitations of the
${\rm NL\sigma M}$. With increasing temperature, the excitation
broadens as may be seen at $T=9\;\mathrm{K}$, where the line-shape
can be described equally well by a Gaussian or Lorentzian
line-shape.\par

Another spectral form that describes damped excitations is the
damped harmonic oscillator (DHO) function which can be written as:
\begin{equation}
        S(\bbox{Q},\omega) = \frac{A \cdot \left( n(\omega) + 1
        \right) 4\Gamma \epsilon(\bbox{Q}) \omega}{(\omega^2 -
        \epsilon_{\rm DHO} (\bbox{Q})^2)^2 + 4 \Gamma^2 \omega^2}
        \label{dho form}
\end{equation}$\epsilon_{\rm DHO}(\bbox{Q})$ is the DHO energy and the
other symbols are as in Eq.~\ref{Lorentzian_form}. The DHO
function is the same function as the antisymmetrized Lorentzian
defined in Eq.~\ref{Lorentzian_form} if
\begin{equation}
        \epsilon_{\rm DHO} (\bbox{Q})^2 =
        \epsilon(\bbox{Q})^2+\Gamma^2\, .
        \label{lorz-dho-energy}
\end{equation}For the same spectral distribution, the DHO
energy $\epsilon_{\rm DHO}(\bbox{Q})$ is therefore larger than the
Lorentzian energy $\epsilon(\bbox{Q})$.\par

\begin{figure}
\begin{center}
  \includegraphics[height=5.5cm,bbllx=105,bblly=330,bburx=500,
  bbury=580,angle=0,clip=]{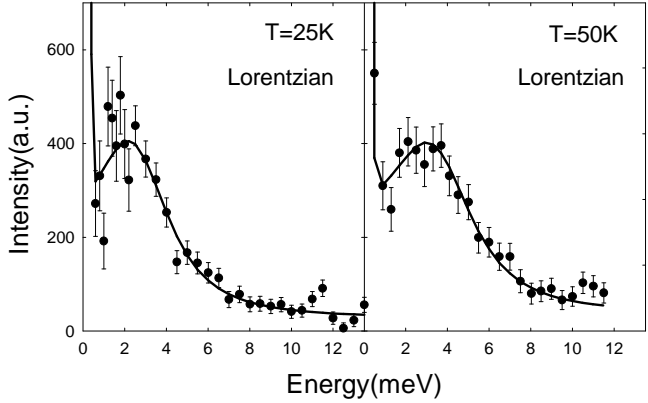}
  \vspace{0.3cm}
  \caption{Neutron scattering intensity at $\bbox{Q}=(0.81,0.81,1)$
  at $T=25$ and $50\;\mathrm{K}$ measured using RITA. The data
  is shown as a function of neutron energy transfer. The solid
  lines correspond to a Lorentzian cross-section convoluted with
  the resolution ellipsoid and fitted to the data.}
  \label{Risoe-scans+fits}
\end{center}
\end{figure}

In Fig.~\ref{Risoe-scans+fits} we show that neutron scattering at
$\bbox{Q}=(0.81,0.81,1)$ at $T=25$ and $50\;\mathrm{K}$ is well
described by a Lorentzian cross-section. The background was
assumed to be constant and a Gaussian was used to account for the
incoherent elastic scattering. The cross-section
(Eq.~\ref{Lorentzian_form}), folded with the resolution
function,\cite{Cooper_Nathans} was fitted to the data to derive
$\epsilon(\bbox{Q})$, $\Gamma$ and $A$.\par

\begin{figure}
\begin{center}
  \includegraphics[height=6.6cm,bbllx=80,bblly=265,bburx=478,
  bbury=584,angle=0,clip=]{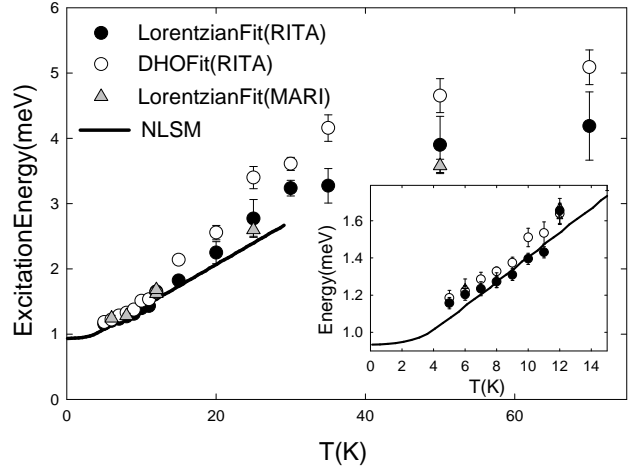}
  \vspace{0.3cm}
  \caption{Excitation energy $\epsilon(\bbox{Q})$ for
  $\bbox{Q}=(0.81,0.81,1))$ as a function of temperature between
  $T=5$ and $70\;\mathrm{K}$. The energies were obtained from fits
  to a resolution-folded Lorentzian or damped harmonic oscillator
  function. The solid line shows the upward renormalization of the
  gap energy obtained from the ${\rm NL\sigma M}$.\protect\cite{Jolicoeur_Golinelli}
  The inset shows the low-temperature range where ${\rm NL\sigma M}$
  is valid.}
  \label{gap-lorz-dho}
\end{center}
\end{figure}

\begin{figure}
\begin{center}
  \includegraphics[height=6.6cm,bbllx=85,bblly=265,bburx=478,
  bbury=584,angle=0,clip=]{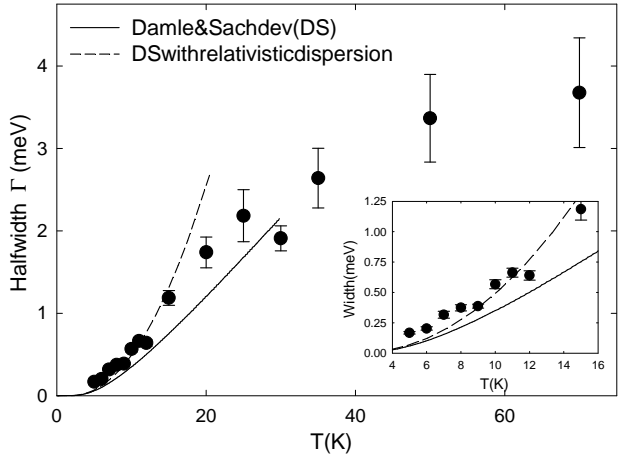}
  \vspace{0.3cm}
  \caption{Half width $\Gamma$, corrected for resolution, for
  $\bbox{Q}=(0.81,0.81,1)$ between $T=5$ and $70\;\mathrm{K}$.
  The widths were obtained from fits to a Lorentzian or a DHO
  function and averaging the widths. The solid line is the
  prediction of Damle and Sachdev. The dashed line is our
  modification of their model to include the relativistic
  dispersion of the ${\rm NL\sigma M}$. The inset shows the
  low-temperature range up to $T=15\;\mathrm{K}$
  where the ${\rm NL\sigma M}$ is valid.}
  \label{width-lorz-dho}
\end{center}
\end{figure}

The excitation energy $\epsilon(\bbox{Q})$ and width $\Gamma$ are
shown in Figs.~\ref{gap-lorz-dho} and \ref{width-lorz-dho} as a
function of temperature $T$. The Haldane gap energy, which at
$T=5\;\mathrm{K}$ is $1.2\;\mathrm{meV}$, increases rapidly
between $T=12$ and $30\;\mathrm{K}$ as the excitation develops
into a broad resonant feature. At about $30\;\mathrm{K}$, a
temperature of the order of the exchange coupling $J$ along the
chain, there is a cross-over to a qualitatively different behavior
and for $T>30\;\mathrm{K}$ the energy increases only slowly with
increasing temperature.\par

The observed gap energies are compared in Fig.~\ref{gap-lorz-dho}
(solid line) with the theory of ${\rm Jolic\oe ur}$ \textit{et
al.}\cite{Jolicoeur_Golinelli} based on the ${\rm NL\sigma M}$,
calculated with the $T=0\;\mathrm{K}$ gap energy of
$\Delta_0=10.6\;\mathrm{K}$ we obtain from the exchange constant
as $\Delta_0=0.406J$. The gap equation of ${\rm
NL\sigma M}$ is a statement that
\begin{equation}
    g\int_0^{\infty} dq \frac{2n(\epsilon_q)+1}{\epsilon_q}=1
    \label{Eq_nlsm}
\end{equation}is a constant in temperature where
\begin{equation}
    \epsilon(q)=\sqrt{\Delta(T)^2+v_s^2 q^2}\, .
    \label{Eq_disp_nlsm}
\end{equation} This
implicit equation for $\Delta(T)$ can be used to calculate the gap
relative to $\Delta_0$ which we take from experiment and use to
obtain $g$ at $T=0\;\mathrm{K}$. For $T<\Delta=11\;\mathrm{K}$
(see inset), where the theory is expected to be valid, the
observed energy matches the theoretical prediction except below
$7\;\mathrm{K}$ where the gap energy is higher. The good agreement
for $7\;\mathrm{K}<T<11\;\mathrm{K}$ shows that the ${\rm NL\sigma
M}$ gives a good description of the low-temperature
quantum-disordered phase for $T<\Delta$. The disagreement below
$T<7\;\mathrm{K}$ is likely due to 3D
interactions,\cite{Kenzelmann_CsNiCl3_continuum_long} These give
rise to low-energy modes as $T_N$ is approached whose thermal
population will drive up the gap energy in the ${\rm NL\sigma M}$
equation (Eq.~\ref{Eq_nlsm}).\par

The measured widths $\Gamma$ are shown in
Fig.~\ref{width-lorz-dho}. The width is always smaller than the
excitation energy up to at least $T=70\;\mathrm{K}$ because the
upward renormalization of the excitation energy is fast enough to
overcome the increasing damping. Damle and Sachdev
\cite{Damle_Sachdev} have developed a theory of the damping of
excitations in gapped spin systems which is valid as long $T <
\Delta$. In ${\rm CsNiCl_3}$ this temperature range is limited
because $\Delta=0.41J=11\mathrm{K}$ and 3D effects are important
close to the ordering temperature $T_N$. As shown in
Fig.~\ref{width-lorz-dho} their theory, which approximates the
${\rm NL\sigma M}$ dispersion by a classical form, $\Delta+cq^2$,
predicts considerably longer excitation life-times than those
observed experimentally. However, we find that Damle and Sachdev's
theory agrees well with the measured excitation width between
$T=9$ and $12\;\mathrm{K}$ if we replace the classical dispersion
of the elementary excitation particles by the more accurate
relativistic dispersion of the ${\rm NL\sigma M}$
(Eq.~\ref{Eq_disp_nlsm}). Details of our calculations are given in
Section \ref{section_discussion}A. Below $T=8\;\mathrm{K}$, the
observed width is consistently higher than that predicted by the
theory. Probably this is due to the 3D spin correlations which are
most important close to $T_N$. We expect that the exponentially
greater thermal population of the low-lying 3D gap modes will
shorten the collision life time relative to that for the gap of an
isolated chain.\par

Above about $T=20\;\mathrm{K} \sim J$ the temperature dependence
of the excitation width decreases, similar to that of the energy.
This behavior probably arises because the lattice momentum cut-off
is not included in the theoretical model. The cut-off is infinite
in Ref.~\cite{Damle_Sachdev}, and in our relativistic calucation
it was limit to $q=\pi$ (see Section VA). An extension of the
theory to higher temperatures would be valuable.\par

\begin{figure}
\begin{center}
\includegraphics[height=6.2cm,bbllx=78,bblly=25,bburx=511,
  bbury=341,angle=0,clip=]{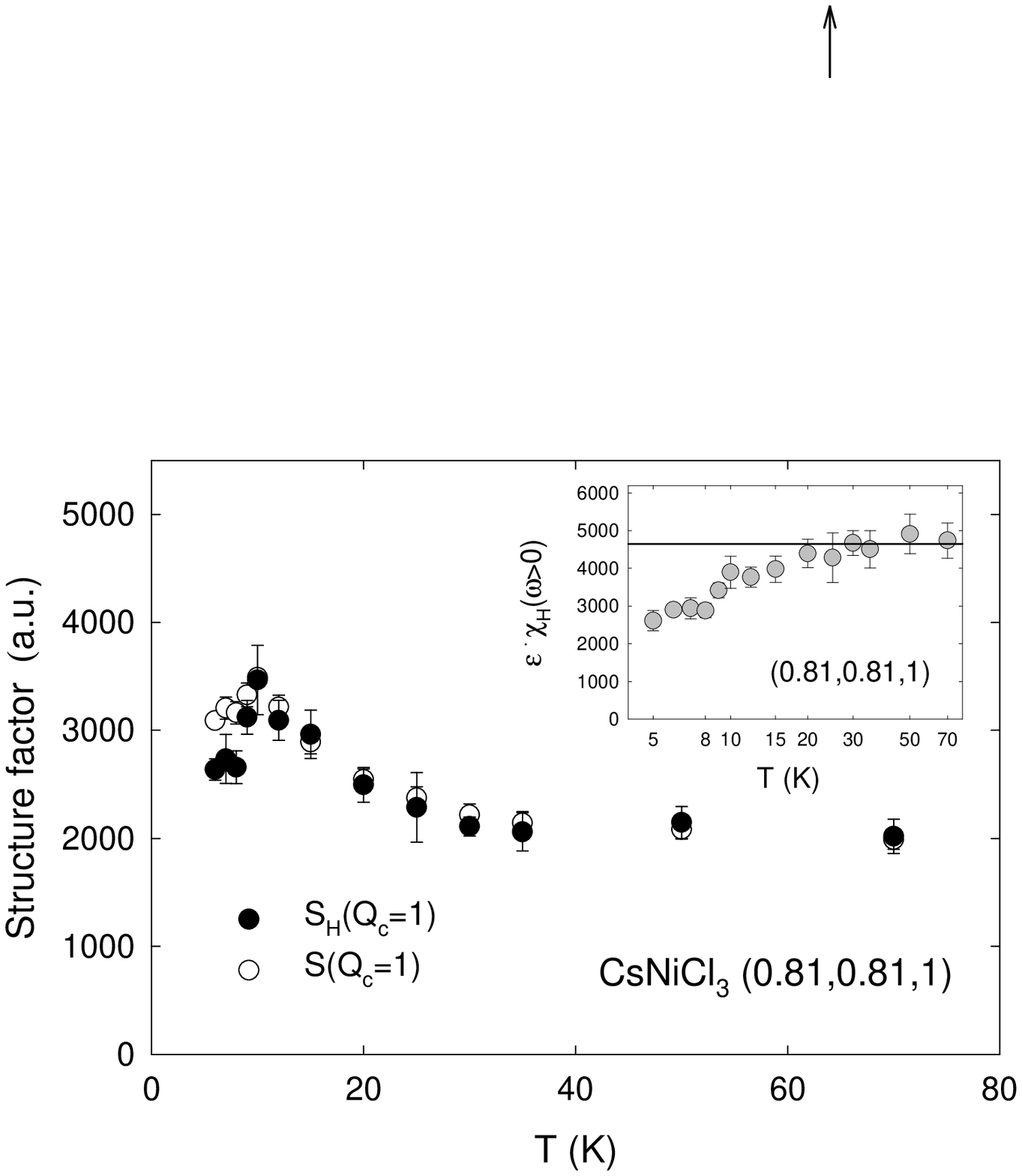}
  \vspace{0.3cm}
  \caption{Structure factor at the 1D point $\bbox{Q}=(0.81,0.81,1)$
  as a function of temperatures. The solid circles show the
  integrated intensity of the Haldane excitation $S_{\rm H}(\bbox{Q})$.
  The open circles represent $S(\bbox{Q})$ determined from the entire
  neutron scattering intensity. Inset: The susceptibility of the Haldane
  excitations for positive energy transfer, defined as
  $\chi_H(\bbox{Q})=\int_0^{\infty}d\omega S_{\rm
  H}(\bbox{Q},\omega)/(n(\epsilon(\bbox{Q}))+1)$, multiplied with the excitation
  energy $\epsilon({\bbox{Q}})$ shows a transition to
  a qualitatively different behavior at $\sim15\;\mathrm{K}$.}
  \label{Risoe-intensity-T}
\end{center}
\end{figure}

Fig.~\ref{Risoe-intensity-T} shows the integrated intensity of the
Haldane excitation, $S_{\rm H}(\bbox{Q})$, for
$\bbox{Q}=(0.81,0.81,1)$, the 1D point, as a function of
temperature (solid circles). The $S_{\rm H}$ was determined from
the observed neutron scattering intensity by integrating the area
under the Lorentzian cross-section for positive and negative
energies. It includes all the scattering except below $\sim
12\;\mathrm{K}$ where the spectrum contains a high energy
continuum of $12(2)\%$ in
weight.\cite{Kenzelmann_CsNiCl3_continuum} $S_{\rm H}$ increases
with decreasing temperature as the antiferromagnetic fluctuations
become stronger and the spectral weight is displaced to lower
energy. The increase between $T=50$ and $20\;\mathrm{K}$ is
associated with the growth of the 1D correlation length. However,
at $10\;\mathrm{K}$ the temperature dependence of $S_{\rm H}$ has
a peak and then decreases with decreasing temperature. This is
correlated with the hold-up in the gap decrease noted for
Fig.~\ref{gap-lorz-dho}.\par

The structure factor $S(\bbox{Q})$ for $\bbox{Q}=(0.81,0.81,1)$,
which includes the continuum, has a lesser decrease at low
temperatures (Fig.~\ref{Risoe-intensity-T}). $S(\bbox{Q})$ was
determined by numerical integration of the observed neutron
spectrum at positive energy transfers and adding intensity at
negative energy transfer calculated using the detailed balance
relation. $S(\bbox{Q})$ and $S_{\rm H}(\bbox{Q})$ are identical at
high temperatures, but for $T < 10\;\mathrm{K}\;$, where the
continuum scattering was observed, $S(\bbox{Q})$ is larger than
$S_{\rm H}(\bbox{Q})$ and approximately constant. The increase in
$S(\bbox{Q})$ on cooling is substantially less than that
calculated for isolated chains.\cite{Kim}

To test whether the spectral weight scales with
$1/\epsilon(\bbox{Q})$ we divided $S_H$ by the population factor
$n(\epsilon(\bbox{Q}))+1$ to obtain the susceptibility for
positive energy transfer $\chi_H$ and multiplied it with the
excitation energy $\epsilon(\bbox{Q})$. This approximately removes
the thermal factor to a degree that improves as the temperature
and width decrease. Our scaled function
(Fig.~\ref{Risoe-intensity-T})is a constant for
$T>15\;\mathrm{K}$, showing that the intensity scales with
$1/\epsilon(\bbox{Q})$. For $T<15\;\mathrm{K}$, however, $\chi_H$
is lower than expected from a $1/\epsilon(\bbox{Q})$-dependence of
the intensity, thus the spectral weight grows more slowly than
$1/\epsilon(\bbox{Q})$.\par

The reason for the behavior below $T=10\;\mathrm{K}$ is at present
unclear. In Kenzelmann \textit{et
al.}\cite{Kenzelmann_CsNiCl3_continuum} we have argued that the
continuum scattering is generated by the interchain coupling which
has a strong effect on the dispersion for temperatures close to
the ordering temperature $T_N$ despite its small size. As will be
shown in a forthcoming publication \cite{Kenzelmann_Santini} the
interchain coupling has a strong effect on the excitation
intensity for all temperatures $T<50\;\mathrm{K}$ and there is
considerably less magnetic intensity in $S_{\rm H}(\bbox{Q})$ as
well as in $S(\bbox{Q})$ near $Q_c=1$ than in uncoupled chains. We
may also imagine as $T_N$ is approached that spectral weight is
transferred from the 1D point to the 3D ordering wave-vector.\par

\subsection{The Dispersion Relation}
The spectrum of excitations of ${\rm CsNiCl_{3}}$ was measured for
$T=12$, $25$ and $50\;\mathrm{K}$ over a wide range of wave-vector
transfers using the MARI spectrometer with the incident beam
perpendicular to the c-axis. The neutron scattering was measured
for energy transfers of up to $75\%$ of the incident energy. For a
1D magnet, the scattering $S(\bbox{Q},\omega)$ was obtained as a
function of $Q_c$ and $\omega$, where $Q_c$ is the component of
the wave-vector transfer along the chain, while the in-plane
wave-vector $Q_a$ is a known function of $Q_c$ and of the energy
transfer.\par

\begin{figure}
\begin{center}
  \includegraphics[height=10cm,bbllx=90,bblly=185,bburx=480,
  bbury=705,angle=0,clip=]{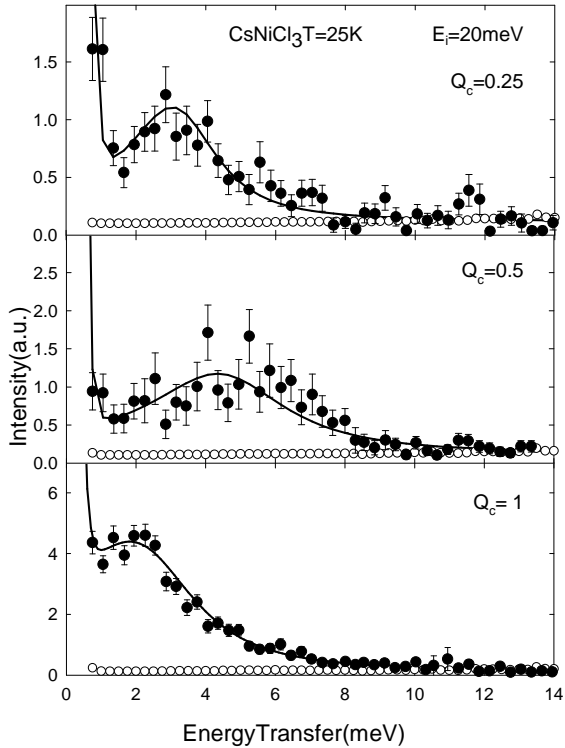}
  \vspace{0.3cm}
  \caption{Neutron scattering spectra at $T=25\;\mathrm{K}$ for
  three different $Q_{\rm c}$. Open circles show the background as
  explained in the text. The solid line is the antisymmetrized
  Lorentzian weighted with the Bose factor.}
  \label{scans-25K}
\end{center}
\end{figure}

\begin{figure}
\begin{center}
  \includegraphics[height=10cm,bbllx=90,bblly=185,bburx=480,
  bbury=705,angle=0,clip=]{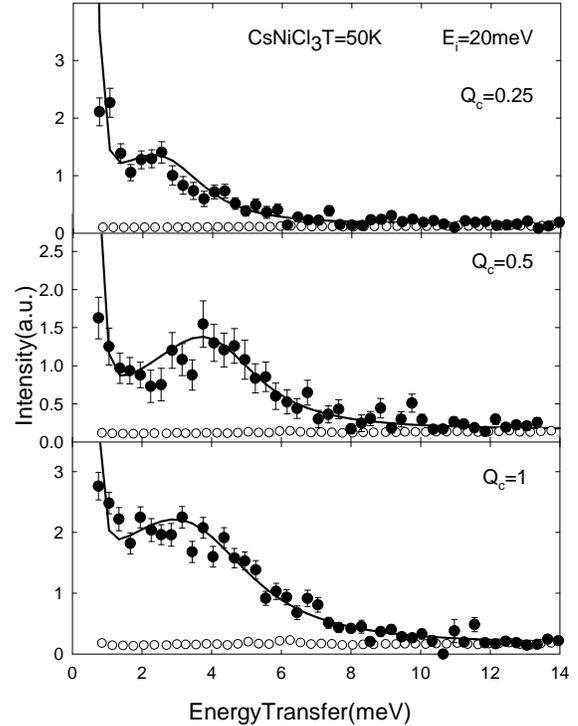}
  \vspace{0.3cm}
  \caption{Neutron scattering spectra at $T=50\;\mathrm{K}$ for
  three different $Q_{\rm c}$. Open circles show the background as
  explained in the text. The solid line is the antisymmetrized
  Lorentzian weighted with the Bose factor.}
  \label{scans-50K}
\end{center}
\end{figure}

For measurements below $T \leq 25\;\mathrm{K}$, where the
excitations may exhibit a dispersion perpendicular to the chain
direction, only data only from the central detector banks were
taken to limit the out-of-plane wave-vectors sampled in the
experiment. Figs.~\ref{scans-25K} and \ref{scans-50K} show the
observed neutron scattering intensity for three wave-vector
transfers along the chain at $T=25$ and $50\;\mathrm{K}$. With
increasing temperature, the intensity for $Q_c=1$ decreases, moves
to higher energies and broadens, as discussed in the previous
section.\par

\begin{figure}
\begin{center}
  \includegraphics[height=8cm,bbllx=80,bblly=310,bburx=490,
  bbury=715,angle=0,clip=]{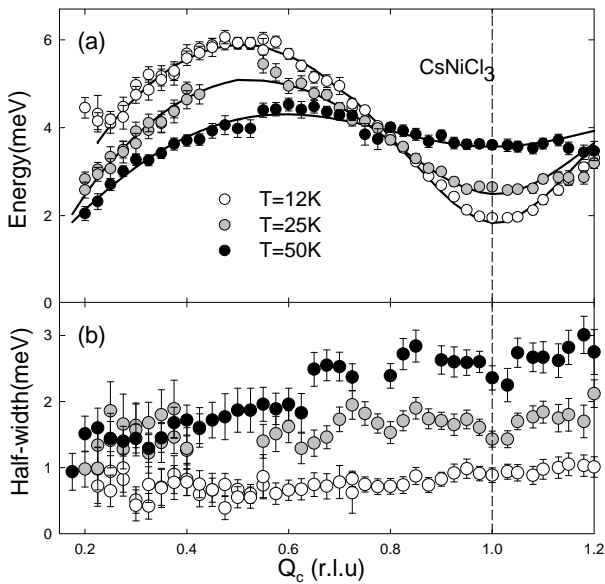}
  \vspace{0.3cm}
  \caption{(a) Excitation energy $\epsilon(Q_c)$ as
  a function of wave-vector and temperature. The lines are
  fits described in the text. (b) Lorentzian half width
  $\Gamma$.}
  \label{dispersion-T}
\end{center}
\end{figure}

Our previous study has shown that the excitation at $Q_c=1$
remained resonant up to $70\;\mathrm{K}$.
\cite{Kenzelmann_CsNiCl3_gap} Figs.~\ref{scans-25K} and
\ref{scans-50K} further show that the excitations remain resonant
throughout the zone and that they are weakly dispersive even at
$50\;\mathrm{K}$. The scattering was again fitted to the
Lorentzian of Eq.~\ref{Lorentzian_form} but now taking the
line-shape of the quasi-elastic peak for the energy resolution.
Least-square fits gave the parameters of
Eq.~\ref{Lorentzian_form}, $A$, $\epsilon(\bbox{Q})$ and $\Gamma$
(see in Fig.~\ref{dispersion-T}). Data with $Q_c>1.3$ was not
analyzed because it was contaminated with phonon scattering.\par

The non-magnetic background was estimated from the scattering
observed in detectors at high scattering angles where the
wave-vector transfer $|\mathbf{Q}|$ is large, the magnetic form
factor is low and the spectrum is dominated by the phonon
scattering which is proportional to $|\mathbf{Q}|^2$. The
non-magnetic background for each scan was obtained by scaling the
intensity according to $|\mathbf{Q}|^2$ and accounting for
background scattering that is independent of $|\mathbf{Q}|$. The
latter was estimated from the energy gain side of the spectrum at
low temperatures. The background is shown in Figs.~\ref{scans-25K}
and \ref{scans-50K} by open circles. Assuming that the whole
intensity at high scattering angle scales only with
$|\mathbf{Q}|^2$, or that part of the scattering scales with
$|\mathbf{Q}|^4$ leads to even lower background estimates for the
low-angle data. The results show that the background is weakly
dependent on energy transfer and therefore may be accurately
subtracted.\par

The dispersion of the excitations decreases with increasing
temperature as shown in Fig.~\ref{dispersion-T}a. While the
excitation energy increases with increasing temperature for
antiferromagnetic fluctuations, it decreases for $Q_c=0.25$ and
$0.5$. Due to the lack of a theoretical model for the dynamic
structure factor for the whole magnetic zone the dispersion was
phenomenologically described by
\begin{eqnarray}
     &\omega^2(Q_c,T) = &\nonumber  \\ & \Delta^2(T) + v_{\rm s}^2(T)
     \sin^2(\pi Q_c)+\alpha^2(T) \cos^2(\frac{\pi Q_c}{2})
     \, .&
     \label{1D-disp-temp}
\end{eqnarray}Here $\Delta(T)$ is the Haldane gap at $Q_c=1$,
$v_s(T)$ is the spin velocity determining the increase of the
excitation energy as $|Q_c-1|$ increases, and $\alpha(T)$ takes
account of the two-particle peak at low $T$ as $Q_c \rightarrow 0$
($\alpha^2=3$), but in practice $\alpha$ is used as a free
parameter to account for the asymmetry about $Q_c=0.5$. For
$T\leq25\;\mathrm{K}$ 3D effects were taken into account using a
RPA approximation which introduces a small dispersion
perpendicular to the chain axis.\cite{Tun90} The dispersion then
reads
\begin{eqnarray}
     &\omega^{2}({\mathbf{Q}}) = \omega(Q_c)
     \cdot \left[ \omega(Q_c)+ 2 \cdot J'(Q_a,Q_b,0)
     \cdot S({\mathbf{Q}}) \right]\, .&
     \label{RPA_disp-temp}
\end{eqnarray}where $J'(Q_a,Q_b,0)$ is the Fourier transform of the
interchain coupling and $S({\mathbf{Q}})=\int
S({\mathbf{Q}},\omega)d\omega$ is the structure factor. The
dispersion perpendicular to the chain is most pronounced at
$Q_c=1$ where $S(\mathbf{Q})$ is maximum and is small at
$Q_c=0.5$.\par

\begin{figure}
\begin{center}
  \includegraphics[height=6.5cm,bbllx=65,bblly=325,bburx=510,
  bbury=675,angle=0,clip=]{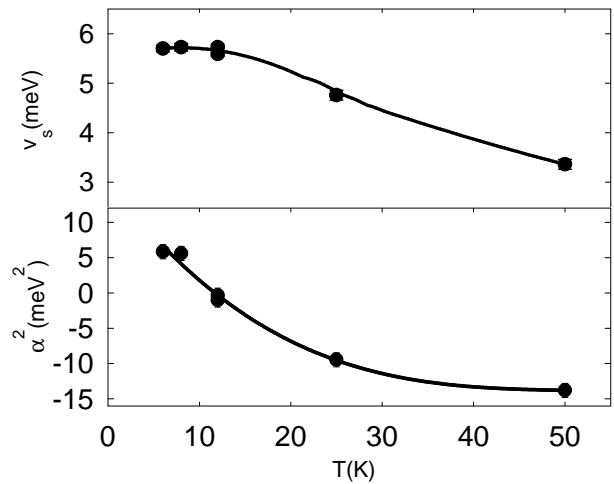}
  \vspace{0.3cm}
  \caption{The upper panel shows the spin velocity $v_s$
  and the lower panel $\alpha^2$ as a function of temperature.
  The solid lines are guides to the eye.}
  \label{Fig-vs-alpha}
\end{center}
\end{figure}

The Haldane gap or resonance energy $\Delta$, the spin velocity
$v_s$ and $\alpha$ were obtained for the different temperatures by
fitting Eq.~\ref{RPA_disp-temp} to the $\mathbf{Q}$-dependence of
the excitation energies. The results for $\Delta(T)$ are shown in
Figs.~\ref{gap-lorz-dho}a and \ref{dispersion-T} and are
consistent with previous results.\cite{Kenzelmann_CsNiCl3_gap} The
results for $v_s(T)$ and $\alpha^2(T)$ are shown in
Fig.~\ref{Fig-vs-alpha}. The spin velocity $v_s$ is approximately
constant below $15\;\mathrm{K}$ at $5.70(7)\;\mathrm{meV}$ and
above $15\;\mathrm{K}$ it decreases with temperature reaching
$3.4(1)\;\mathrm{meV}$ at $50\;\mathrm{K}$. The asymmetry
parameter $\alpha^2$ is $6\;\mathrm{meV^2}$ at $T=6\;\mathrm{K}$,
decreases rapidly with increasing temperature and becomes negative
above about $12\;\mathrm{K}$. Since it is merely a
phenomenological parameter, this is not unphysical but reflects
the reversal of the asymmetry about $Q_c=0.5$ as the high
temperature limit is approached.\par

The excitations strongly broaden with increasing temperature
(Fig.~\ref{dispersion-T}b). Below $T=25\;\mathrm{K}$, the width is
approximately independent of wave-vector. For $Q_c>1.2$ the width
begins to increase and we have evidence that the scattering there
is contaminated with phonons and so these data were not further
analyzed. For $T=50\;\mathrm{K}$ the excitation width increases
from $Q_c=0.2$ to $Q_c=1$ - an indication of a change in its
$Q_c$-dependence with increasing temperature. We shall see that
this is expected at high temperatures. The widths shown in
Fig.~\ref{dispersion-T}b are slightly smaller relative to those in
Fig.~\ref{width-lorz-dho}, presumably because the resolution width
was overestimated in our analysis.\par

\subsection{Structure factor and correlation length}
The structure factor $S(\bbox{Q})$ was determined numerically from
the measured neutron spectra at positive energy transfer after
subtraction of the flat background. The intensity at negative
energy transfers was estimated using detailed balance. The
structure factor was corrected for the wave-vector dependence of
the magnetic form factor and for the effect of the 3D interactions
by using Eq.~\ref{RPA_disp-temp} to derive the structure factor
for the fluctuations at the 1D point:
\begin{equation}
    S(Q_c)=\frac{\omega(\bbox{Q})}{\omega(Q_c)}S(\bbox{Q})
    \, .
    \label{SRL-3D-temp}
\end{equation}This can be thought of as the structure factor of a
single chain embedded in the system.\par

\begin{figure}
\begin{center}
  \includegraphics[height=6.5cm,bbllx=70,bblly=260,bburx=480,
  bbury=570,angle=0,clip=]{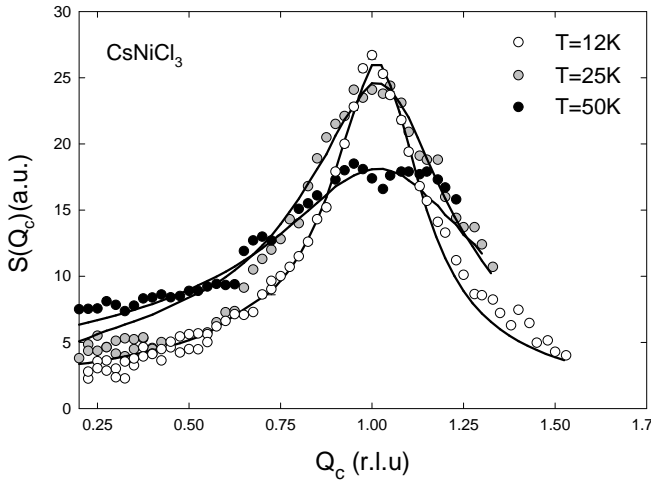}
  \vspace{0.3cm}
  \caption{Structure factor $S(Q_c)$ along $(0.81,0.81,Q_c)$
  for $T=12$, $25$ and $50\;\mathrm{K}$ after an allowance for
  3D effects (Eq.~\protect\ref{SRL-3D-temp}). $S(Q_c)$ was corrected
  for background and magnetic form factor. The solid lines correspond
  to a fit to a square-root Lorentzian
  (Eq.~\protect\ref{SRL-temp}) as explained in the text.}
  \label{intensity-T}
\end{center}
\end{figure}

The effective structure factor for the 1D chain
(Fig.~\ref{intensity-T}) shows strong antiferromagnetic
fluctuations at $Q_c=1$ that persist to $T \sim 2J$. They decline
with temperature while $S(Q_c)$ at $Q_c=0.25$ and $0.5$ increases.
For a 1D magnet with a magnetic excitation spectrum that is
dominated by a single mode, the $Q_c$-dependence of the structure
factor $S^{\alpha \alpha}(Q_c)$, $\alpha=x,\,y,\,z$, can be
expressed as\cite{Hohenberg_Brinkman}

\begin{equation}
    S^{\alpha \alpha}(Q_c) = - \frac{4}{3} \frac{{<\mathcal{H}>}}{L}
    \frac{(1-\cos{\pi Q_c})}{\hbar \omega^{\alpha \alpha}(Q_c)}\,
    .
    \label{SMA-temp}
\end{equation}Here ${<\mathcal{H}>}/L$ is the ground state energy per
spin and $\omega^{\alpha \alpha}(Q_c)$ is the single mode energy.
This approximation is called the single mode approximation (SMA).
Taking the energy from Eq.~\ref{1D-disp-temp}, the structure
factor close to the antiferromagnetic point behaves like a
square-root Lorentzian:
\begin{equation}
    S(Q_c) \propto \frac{\xi}{ \sqrt{ 1+ \pi^2 (Q_c-1)^{2}
    \cdot \xi^{2}}}\, ,
    \label{SRL-temp}
\end{equation}where we expect
\begin{equation}
    \xi=\frac{\sqrt{v_{\rm s}^2+\alpha^2/4}}{\Delta}\; ,
    \label{xi_vs_alpha}
\end{equation}and where $\xi$ is the correlation length. $\xi$ was
determined from $S(Q_c)$ for $0.7<Q_c<1.3$ by fitting
Eq.~\ref{SRL-temp} to the data shown in
Fig.~\ref{intensity-T}.\par

\begin{figure}
\begin{center}
\includegraphics[height=6.5cm,bbllx=75,bblly=250,bburx=480,
  bbury=570,angle=0,clip=]{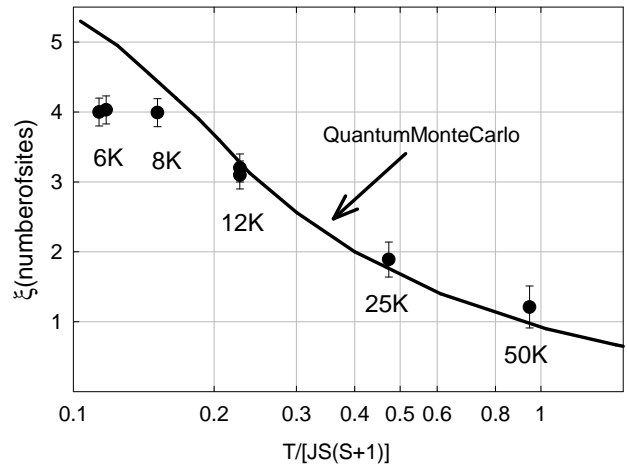}
  \vspace{0.3cm}
  \caption{Correlation length $\xi(T)$ deduced from $S(Q_c)$
  and shown as a function of the reduced temperature
  $T/(JS(S+1))$ on a semi-logarithmic plot. The solid line is
  the prediction of the correlation length obtained from a
  quantum Monte Carlo calculation.\protect\cite{Kim}}
  \label{corr-length}
\end{center}
\end{figure}

The temperature dependence of the correlation length $\xi$ is
shown in Fig.~\ref{corr-length} on a semi-logarithmic plot. At
$50\;\mathrm{K}$ it is only slightly longer than $1$ site which
means that the spatial correlations are extremely short ranged. It
is surprising to have resonant and dispersive modes in a spin
system with such a short correlation length. With decreasing
temperature, the correlation length increases to about 4 sites
below $8\;\mathrm{K}$, as previously
reported.\cite{Kenzelmann_CsNiCl3_continuum_long} These results
can be compared to the correlation length obtained by Kim
\textit{et al.}\cite{Kim} from numerical quantum Monte Carlo
calculations in which effective spin $S$ chains were realized by
ferromagnetically coupling $n=2S$ antiferromagnetic spin chains
with $S=1/2$.\cite{Kim} Our measured correlation length for
temperatures above $12\;\mathrm{K}$ is in excellent agreement with
the numerical results, and only below $12\;\mathrm{K}$ where 3D
correlations are most important is the correlation length shorter
than the numerical prediction for uncoupled spin-1 chains. That it
becomes shorter is associated with the fact that we are at the
non-critical 1D wave-vector.\par

\begin{figure}
\begin{center}
\includegraphics[height=6.5cm,bbllx=75,bblly=250,bburx=480,
  bbury=570,angle=0,clip=]{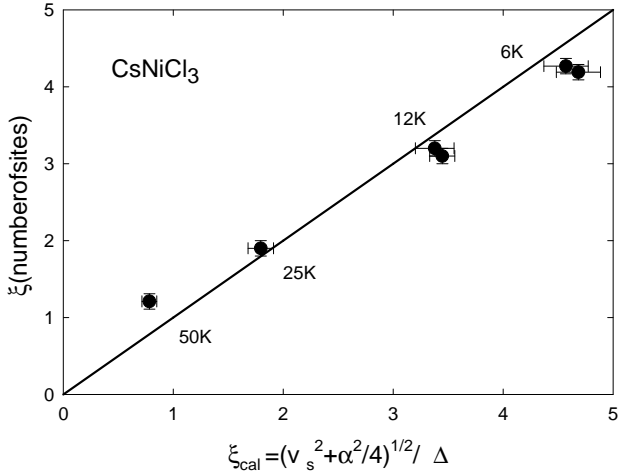}
  \vspace{0.3cm}
  \caption{The horizontal axis represents the correlation length
  $\xi_{\rm cal}$ calculated from the experimentally determined
  $\Delta$, $v_s$ and $\alpha$ for temperatures between $T=6$ and
  $50\;\mathrm{K}$. The vertical axis shows the correlation
  length $\xi$ determined from the wave-vector dependence of the
  structure factor $S(Q_c)$. The solid line is $\xi_{\rm cal}=\xi$.}
  \label{gap-vs-corr}
\end{center}
\end{figure}

For a 1D magnet in which the magnetic response is dominated by
single mode excitations, the SMA (Eq.~\ref{1D-disp-temp}) predicts
that $\xi$ is given by Eq.~\ref{xi_vs_alpha}. In this experiment,
all four quantities were independently determined and the
relationship can be tested experimentally. Fig.~\ref{gap-vs-corr}
compares the correlation length $\xi$ estimated from the
$Q_c$-dependence of the structure factor and the correlation
length $\xi_{\rm calc}$ given by Eq.~\ref{xi_vs_alpha}. The
agreement is excellent for all temperatures. This demonstrates
that SMA is valid up to $T=50\;\mathrm{K}$ and suggests that the
observed excitations can be regarded as short-lived single
particle excitations.\par

The Hohenberg-Brinkman sum-rule\cite{Hohenberg_Brinkman} predicts
that the first energy moment $F(Q)=\int_{\infty}^{\infty} d\omega
\omega S(Q,\omega)$ of a 1D magnet with isotropic nearest-neighbor
exchange $J$ is given by
\begin{equation}\label{Eq_HB}
    F(Q_c) = - \frac{4}{3}\frac{<{\mathcal{H}}>}{L} (1-\cos(\pi Q_c))\, .
\end{equation}

\begin{figure}
\begin{center}
  \includegraphics[height=5.5cm,bbllx=50,bblly=240,bburx=550,
  bbury=550,angle=0,clip=]{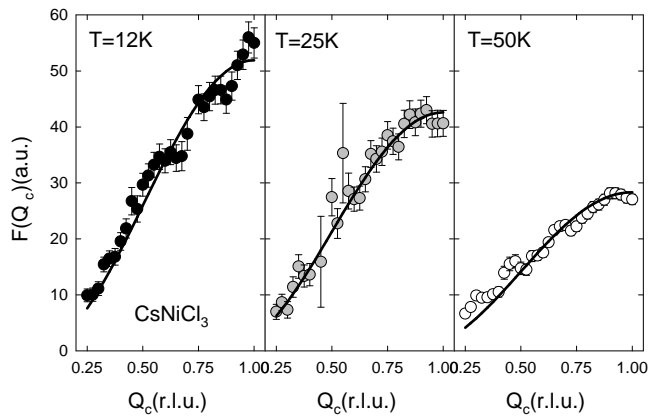}
  \vspace{0.3cm}
  \caption{First energy moment $F(Q_c)$=$\int_{\infty}^{\infty} d\omega\,
  \omega\, S(Q,\omega)$ for $T=12$, $25$ and $50\;\mathrm{K}$.
  The solid line is a fit of the Hohenberg-Brinkman sum rule
  (Eq.~\ref{Eq_HB}) to a parameter proportional to the internal
  energy $<{\mathcal{H}}>$.\protect\cite{Hohenberg_Brinkman}}
  \label{hohenberg-high}
\end{center}
\end{figure}
Here $Q_c$ is the wave-vector transfer along the spin chain and
$<{\mathcal{H}}> = 3 J \sum_i <S^{\alpha}_i S^{\alpha}_{i+1}>$ is
the expectation value of the Hamiltonian.\par

The first energy moment $F(Q_c)$ was determined numerically from
the measured neutron scattering spectrum after subtraction of the
background and accounting for the magnetic form factor, and is
shown in Fig.~\ref{hohenberg-high} for three different
temperatures. The dependence on $Q_c$ of the first moment,
$F(Q_c)$, is in excellent agreement with the sum-rule for all
momenta and temperatures. The fit yields the internal energy
$<{\mathcal{H}}>$ in arbitrary units and its temperature
dependence, and it will be discussed in a forthcoming
publication.\cite{Kenzelmann_Santini} The data for
$<{\mathcal{H}}>$ are shown in Table~\ref{positive_moments}.

\section{High-temperature paramagnetic scattering}
At infinite temperature, there are no correlations between the
spins and the instantaneous structure factor $S(\bbox{Q})$ is a
constant. The first moment of the spectrum vanishes for all
$\bbox{Q}$ because the Bose factor weighs the excitations for
neutron energy gain and loss equally. The second moment, however,
is predicted to vanish at the ferromagnetic momenta and reach a
maximum at antiferromagnetic momenta, irrespective of the spin
quantum number or whether the exchange is antiferromagnetic or
ferromagnetic.\cite{deGennes,Lovesey_book} Chain systems with
nearest-neighbor interactions are ideal systems to test this
behavior, because the wave-vector dependence of the second moment
has a simple form:
\begin{equation}
<\omega^2> \propto J^2 [1-\cos(\pi Q_c)]\, ,
\end{equation}where $Q_c$ is the wave-vector transfer along the
chain.\par

\begin{figure}
\begin{center}
  \includegraphics[height=7.25cm,bbllx=113,bblly=260,bburx=478,
  bbury=565,angle=0,clip=]{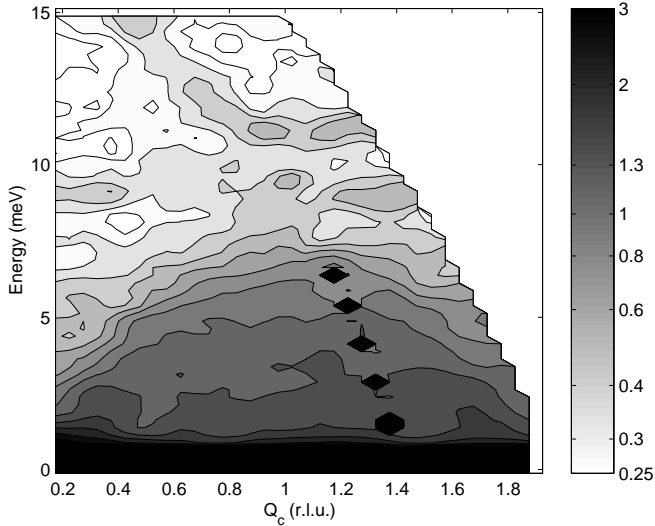}
  \vspace{0.3cm}
  \caption{The intensity measured at $T=200\;\mathrm{K}$ using MARI.
  The data were measured with an incident energy $E_i=20\;\mathrm{meV}$
  and the intensity in all three detector banks was added and
  smoothed. The intensity is indicated by the bar on the right-hand side.
  The black trapezoidal areas are an artifact and a result of a MARI
  detector gap in the $2\Theta$ scattering angle and the data smoothing
  algorithm. The quasi-elastic scattering below $\sim 1\;\mathrm{meV}$ should
  be ignored.}
  \label{MARI-plot-200K}
\end{center}
\end{figure}

The scattering was measured at $T=200\;{\mathrm{K}} \simeq 10J$
for a wide range of wave-vector transfers using the MARI
spectrometer. Fig.~\ref{MARI-plot-200K} shows the scattering
surface projected onto the $(Q_c,\omega)$ plane for neutron energy
loss. Because the effect of the interchain interaction is
negligible at $T=200\;\mathrm{K}$, the data from all the detectors
were used to improve the statistics. Fig.~\ref{MARI-plot-200K}
shows that the scattering consists of broad scattering and that
there is no sign of the resonant excitations seen at lower
temperatures. The broad dome of spectral weight is centered at
$Q_c=1$ for the high energy fluctuations whereas it was located at
$Q_c=0.5$ at low temperatures.\par

\begin{figure}
\begin{center}
  \includegraphics[height=11cm,bbllx=75,bblly=80,bburx=480,
  bbury=650,angle=0,clip=]{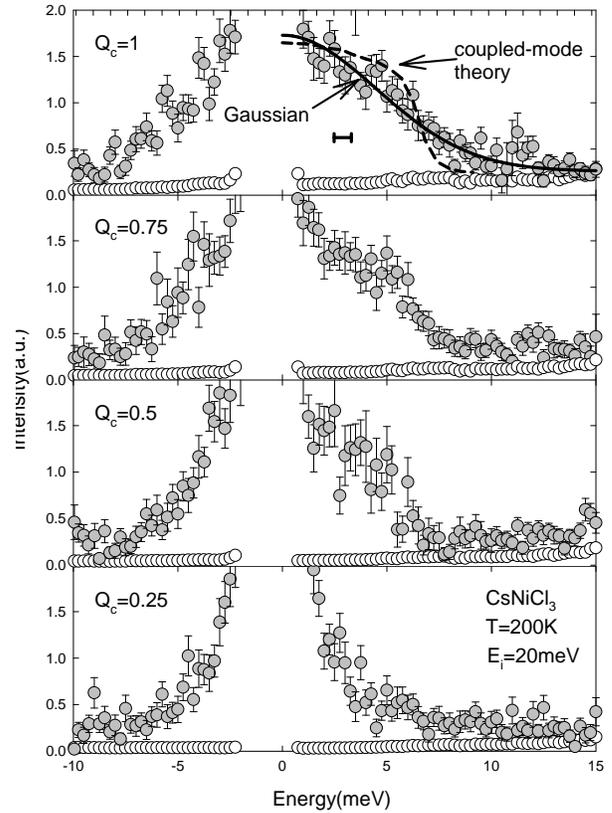}
  \vspace{0.3cm}
  \caption{Neutron spectra at $T=200\;\mathrm{K}$
  for four different $Q_c$. The peak centered at zero energy
  transfer arises from quasi-elastic incoherent scattering from
  the sample and the sample holder. The open circles show the
  background as explained in the text. The dashed and solid lines
  are the prediction of a coupled-mode theory at infinite
  temperature,\protect\cite{Lovesey_Balcar} and the de-Gennes
  Gaussian fitted to the data, respectively. The instrumental
  resolution (FWHM) is shown by a bar.}
  \label{MARI-cuts-200K}
\end{center}
\end{figure}

Fig.~\ref{MARI-cuts-200K} shows constant-$Q_c$ scans for different
wave-vector. The upper bound of the scattering has a maximum of
$10\;\mathrm{meV}$ at $Q_c=1$ and decreases with increasing
$|Q_c-1|$.\par

The scattering intensity observed at the antiferromagnetic point
$Q_c=1$ is not consistent with the prediction of a coupled-mode
theory at infinite temperature,\cite{Lovesey_Balcar} shown as a
dashed line in Fig.~\ref{MARI-cuts-200K}. Instead the scattering
spectrum has the form of a Gaussian as predicted by de
Gennes\cite{deGennes} (solid line). The difference between the
coupled-mode theory and the experiment may be because the
experiment was performed at a finite temperature but more likely
because the theory is unsatisfactory.\par

\begin{figure}
\begin{center}
  \includegraphics[height=7cm,bbllx=70,bblly=275,bburx=480,
  bbury=645,angle=0,clip=]{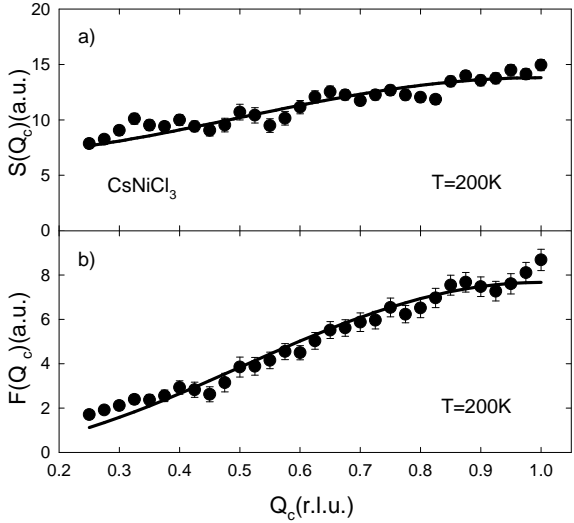}
  \vspace{0.3cm}
  \caption{(a) $S(Q_c)$ as a function of wave-vector transfer
  $Q_c$ at $T=200\;\mathrm{K}$ in the same units as in
  Fig.~\ref{intensity-T}. The solid line is the prediction
  of the high-temperature series expansion
  (Eq.~\ref{Eq_SQ}),\cite{Lovesey_book} normalized to the data.
  (b) Observed first energy moment $F(Q_c)$. The solid line is the
  fit of the Hohenberg-Brinkman sum rule,
  $-\frac{4}{3}\frac{<{\mathcal{H}}>}{L} (1-\cos(\pi Q_c))$.}
  \label{MARI-int-first-200K}
\end{center}
\end{figure}

The structure factor $S(Q_c)$ and the first energy moment $F(Q_c)$
are shown in Fig.~\ref{MARI-int-first-200K} in the same units as
for the lower temperatures (Figs.~\ref{dispersion-T} and
\ref{intensity-T}). They were calculated numerically from the
observed neutron scattering at positive neutron energy transfers
after subtraction of a flat background that matched the scattering
at high energy transfers. The intensity at negative energy
transfers was estimated using detailed balance. $S(Q_c)$ and
$F(Q_c)$ were corrected for the magnetic form factor. $S(Q_c=1)$
is now only about twice $S(Q_c=0.25)$, showing that there are weak
spin correlations, but these are considerably reduced compared to
lower temperatures. The solid line is the prediction of a
high-temperature series expansion (Eq.~\ref{Eq_SQ}) which will be
discussed in more detail in Section \ref{section_discussion}.\par

The $Q_c$-dependence of the first moment $F(Q_c)$ is well
described with the sum rule of Hohenberg and
Brinkman\cite{Hohenberg_Brinkman}, confirming that the scattering
is magnetic and that contamination with phonon scattering is not
significant. The maximum of $F(Q_c)$ is $\sim 6.5$ times lower at
$200\;\mathrm{K}$ than the maximum at $12\;\mathrm{K}$, showing
that although the first moment is not zero, the system is close to
the high-temperature limit.\par

It is clear from Figs.~\ref{MARI-plot-200K} and
\ref{MARI-cuts-200K} that the second moment has a maximum at the
antiferromagnetic point $Q_c=1$ as predicted by the theory of de
Gennes. It will be shown in the section \ref{section_discussion}
that there is quantitative agreement between experiment and theory
if the scattering is expanded to first order in $1/k_B T$.\par

\section{Discussion}\label{section_discussion}
\subsection{Life-time of Haldane excitation for ${\bbox T<\Delta}$}
The excitations of an antiferromagnetic isolated spin-1 chain in
its quantum-disordered phase are well-defined, gapped spin-1
excitations (Haldane excitations) which satisfy a relativistic
dispersion relation, $\epsilon(q) = (\Delta^2 + v_s^2 q^2)^{1/2}$,
where $q=|Q-1|$. At zero temperature where the spin correlation
length is $6$ sites,\cite{Kim} the neutron injects an excitation
in the Haldane band which travels with a group velocity $v_s^2
q/\epsilon(q)$, and its life-time is infinitely long. At a finite
temperature $T \leq \Delta=0.41 J$, thermal activation creates a
dilute gas of Haldane excitations. They limit the life-time of the
injected excitation through collisions, and this gives it a finite
energy.\cite{Damle_Sachdev}\par

The scattering matrix for a two-particle collision has a
superuniversal form in 1D which does not depend on any microscopic
parameter, and  for $T\ll\Delta$, it predicts that a two-particle
state ($\{q_1,\sigma_1\},\{q_2,\sigma_2\}$) transforms into a
final state ($\{q_1,\sigma_2\},\{q_2,\sigma_1\}$), where $\sigma$
is the spin quantum number, thus basically swapping the
spin.\cite{Damle_Sachdev} Alternatively in 1D we may think that
the particle created by the neutron, $\{q_1,\sigma_1\}$, collides
with one of the gap particles $\{q_2,\sigma_2\}$ and, like
billiard balls, the momentum of the first is transmitted to the
second, to give $\{q_2,\sigma_1\}$ and $\{q_1,\sigma_2\}$. This
result holds for any form of 1D dispersion relation. Damle and
Sachdev\cite{Damle_Sachdev} calculated the temperature dependence
of the line-shape and width by approximating the relativistic
dispersion with a classical dispersion, $\epsilon(q) = \Delta +
\frac{v_s^2 q^2}{2\Delta}$, and when $T\ll\Delta$, the dynamic
structure factor has a simple scaling form
\begin{equation}
S(Q,\omega) = \frac{A c}{\gamma \Delta} \Phi \left(
\frac{\omega-\epsilon(q)} {\gamma}\right) \label{structure_factor}
\end{equation}
where $\gamma$ is the inverse of the time between collisions with
other excitations and $A$ is a scaling constant. The scaling
function $\Phi(z)$ was calculated numerically and found to be
close to a Lorentzian form:
\begin{equation}
\Phi(z) = \frac{\pi \alpha}{2(\alpha^2+z^2)}\, , \label{function}
\end{equation}with $\alpha=0.71$. Thus the
theory predicts that the observed half-width is $\Gamma=\alpha
\cdot \gamma$. The inverse collision time is proportional to the
density of the excitations times their root mean square velocity,
namely
\begin{equation}
\gamma(T) = \frac{3 k_{B}T}{\sqrt{\pi}}
\exp\left(-\frac{\Delta}{k_{B}T}\right)\, , \label{theorie_width}
\end{equation}the same as for classical gas collisions. The
experimentally observed excitation $\Gamma$ width in ${\rm
CsNiCl_3}$ is appreciably lower than this prediction as shown in
Fig.~\ref{width-lorz-dho} for all
$T<\Delta\simeq11\;\mathrm{K}$.\par

We have modified the calculation of the life-time of the
excitations by starting from the proper relativistic dispersion
instead of the classical approximation used by Damle and
Sachdev.\cite{Damle_Sachdev} We have also applied a finite lattice
momentum cut off. The density of the quasi-particle gas is then
given by
\begin{equation}
\rho_{\rm rel}(T) = 3 \int_{-\pi}^{\pi} \frac{dq}{2\pi}
\exp\left(-\frac{\sqrt{\Delta^2+v_s^2q^2}}{k_B T}\right)
\end{equation}
The mean square velocity of the quasi-particles is given by
\begin{equation}
v_{\rm rel}^2(T) = \frac{3}{\rho_{\rm rel}(T)} \int_{-\pi}^{\pi}
\frac{dq}{2\pi} (v_s^2 q)^2
\exp\left(-\frac{\sqrt{\Delta^2+v_s^2q^2}}{k_B T}\right)
\end{equation}
These two integrals are not analytically solvable but were
calculated numerically to obtain $\gamma_{\rm rel}(T)=v_{\rm
rel}(T) \rho_{\rm rel}(T)$. $\gamma_{\rm rel}$ is larger than
$\gamma$ because the excitation energy of relativistic particles
is lower at large momenta than for classical particles with the
same $\Delta$ and $v_s$. At the same temperature more excitations
are then populated so that collisions are more frequent.\par

$\Gamma_{\rm rel}=\alpha\cdot\gamma_{\rm rel}$ is shown in
Fig.~\ref{width-lorz-dho} as dashed line, and in the expected
range of applicability the agreement between this calculation and
the experiment is good for $T>8\;\mathrm{K}$. This agreement is
evidence for the validity of the model by Damle and Sachdev that
describes an antiferromagnetic spin-1 chain at $T<\Delta$ as a
dilute gas of Haldane excitations.\cite{Damle_Sachdev} Our
improved calculation merely helps the theory to remain valid to
the relatively high temperatures of our experiment. Below
$T<8\;\mathrm{K}$ the observed excitation widths are higher than
the calculations, most probably due to the interchain spin
correlations which soften the Haldane gap near the ordering
wave-vector $(0.33,0.33,1)$, so that there is an increased thermal
population of excitations.\par
\subsection{Temperature dependence of solitonic excitations in quantum-disordered phase}
\begin{figure}
\begin{center}
\includegraphics[height=5cm,bbllx=23,bblly=250,bburx=550,
  bbury=570,angle=0,clip=]{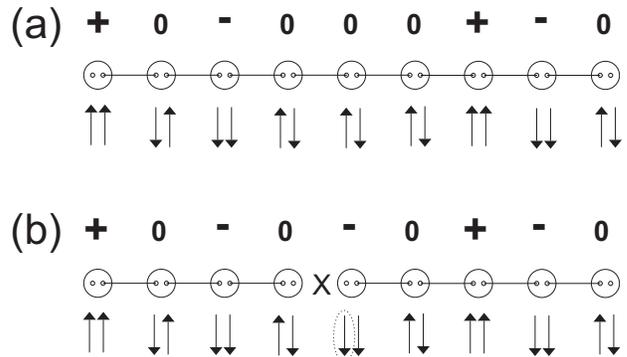}
  \vspace{0.3cm}
  \caption{(a) VBS ground state where the spin-1 (large circles) can
  be pictured as two symmetrized spin-1/2 (small circles within the
  large circles). Two spin-1/2 from neighboring sites form singlet bonds
  (straight lines between two spin-1/2). The picture shows one
  possible spin-1/2 configuration in the ground state and the
  corresponding spin-1 configuration showing the hidden spin string
  order. (b) VBS state after a neutron has transferred $S^z=-1$ to a spin-1,
  creating a symmetrized triplet state between two spin-1/2 on
  neighboring sites (the cross between two spin-1/2). This breaks the
  hidden spin-1 order.}
  \label{Fig_hidden order}
\end{center}
\end{figure}

An antiferromagnetic spin-1 Heisenberg chain does not order at any
finite temperature and the spin-spin correlations decay
exponentially with distance. However, there is a form of hidden
spin order given by a non-local string order parameter first
proposed by Den Nijs and Rommelse.\cite{Den_Nijs_Rommelse} It is
defined as
\begin{equation}
{\mathcal{O}}_{\rm string}^{\alpha}({\mathcal{H}}) = {\rm
lim}_{|i-j|\rightarrow \infty}
<\sigma_{ij}^{\alpha}>_{\mathcal{H}}\, ,
\end{equation}where
\begin{equation}
\sigma_{ij}^{\alpha}=-S_i^{\alpha} \exp\left(i\pi
\sum_{l=i+1}^{j-1}S_l^{\alpha}\right)S_j^{\alpha}\, .
\end{equation}Kennedy and Tasaki\cite{Kennedy_Tasaki} showed the
hidden spin order can be regarded as consequences of the breaking
of the $Z_2 \times Z_2$ symmetry. It was shown that
${\mathcal{O}}_{\rm string}^{\alpha} ({\mathcal{H}})=0.374$ for an
antiferromagnetic chain with nearest-neighbor Heisenberg
interactions.\cite{Girvin_Arovas,Hatsugai_Kohmoto,White_Huse} The
hidden spin order can be pictured in the frame work of the valence
solid bond (VBS),\cite{Affleck} where the $S=1$ spins are obtained
by symmetrizing two $S=1/2$ spins on one site, and each of the
$S=1/2$ spins forms a coherent singlet state with another $S=1/2$
spin from a neighboring site. A possible VBS spin configuration
for the ground state is shown in Fig.~\ref{Fig_hidden order}(a)
together with its configuration in the spin-1 picture. The $S_z=1$
and $S_z=-1$ states alternate with a random number of $S_z=0$
states in between, and the ground state can be viewed as diluted
spin-1 ${\rm N\acute{e}el}$ order, which is the string order
parameter defined above. In VBS ${\mathcal{O}}_{\rm
string}^{\alpha}({\mathcal{H}})=\frac{4}{9}$. This value is higher
than that for the Heisenberg model because quantum fluctuations
reduce the long-range hidden order in the latter.\par

A scattered neutron excites a spin-1/2 pair into a symmetric
triplet state, as shown in Fig.~\ref{Fig_hidden order}(b). The
corresponding spin configuration in the spin-1 picture shows that
the dilute spin-1 ${\rm N\acute{e}el}$ order is now broken. ${\rm
F\acute{a}th}$ and ${\rm S\acute{o}lyom}$ have shown that such a
triplet states are moving hidden domain walls and are one-soliton
excitations of the string-ordered ground
state.\cite{Fath_Solyom}\par

At finite temperature, the hidden non-local string order is
destroyed because spin flips are allowed through thermal
population of the Haldane band that change the instantaneous
magnetization of the chain. Yamamoto and Miyashita
\cite{Yamamoto_Miyashita} introduced a local (short-range) hidden
order parameter
\begin{equation}
{\mathcal{O}}_{\rm SR}=\frac{N(+-)+N(-+)-N(++)-N(--)}
{N(+-)+N(-+)+N(++)+N(--)}
\end{equation}where $N(\pm\mp)$ and $N(\pm\pm)$ are the total number
of nearest spin pairs of $(\pm\mp)$ and $(\mp\mp)$ type,
respectively. The operator is applied to a spin wave-function, for
example $(..-0+-0+0-0+-0+0-..)$, where all spin projection $0$ are
omitted, leaving $(..-+-+-+-+-..)$. The expectation value
$<{\mathcal{O}}_{\rm SR}>$ is $1$ in the absence of $(++)$ and
$(--)$ pairs which can regarded as domain walls.\par

While the non-local string order collapses at any finite
temperature, the local string order decreases only slowly with
increasing temperature,\cite{Yamamoto_Miyashita} and reduces only
by a factor of $2.7$ from $T=0$ to $50\;\mathrm{K}$. We note that
$S(Q_c=1)$ grows by $1.5$ between $50\;\mathrm{K}$ and
$12\;\mathrm{K}$ and is not too much less than the factor of $2$
growth of the local string order.\cite{Yamamoto_Miyashita} In the
light of the persistence of the hidden spin order to high
temperatures it is tempting to interpret the observed excitations
as moving domain walls of the spin string order, and it is less
surprising that they are resonant at such high temperatures. More
theoretical work is needed to clarify this point, and to describe
the energy and the life-time of these high-temperature
solitons.\par

\subsection{High-temperature paramagnetic scattering}
In the high-temperature limit $J\beta=J/k_B T \ll 1$, the second
moment of the dynamic structure factor is given
by\cite{Lovesey_book}
\begin{equation}
<\omega^2>= \frac{-\sum_{i,j}
\exp(i{\bbox{Q}}({\bbox{R}_i}-{\bbox{R}_j}))<[[{\bbox{S}^{\alpha}_{i}},
{\mathcal{H}}],{\bbox{S}^{\alpha}_{j}}]>}
{\beta<{\bbox{S}}^{\alpha}_{\bbox{Q}}
{\bbox{S}}^{\alpha}_{-\bbox{Q}}>} \, .
\end{equation}
The thermal average of the spin-spin correlations is
\begin{equation}
<\bbox{S}^{\alpha}_{i} \bbox{S}^{\delta}_{j}> = \mathrm{Tr}
\left(\exp(-\beta {\mathcal{H}})\bbox{S}^{\alpha}_{i}
\bbox{S}^{\delta}_{j}\right)/\mathrm{Tr}\left(\exp(-\beta
{\mathcal{H}})\right)\, .
\end{equation}To first order in $\beta$,
\begin{equation}
<\bbox{S}^{\alpha}_{i} \bbox{S}^{\delta}_{j}> =
\delta_{\alpha,\delta} \frac{1}{3} S(S+1)
\end{equation}and
\begin{equation}
<{\bbox{S}}^{\alpha}_{i} {\bbox{S}}^{\delta}_{j}> = -
\delta_{\alpha,\delta}\; \beta \left(\frac{1}{3} S(S+1)\right)^2
J({\bbox{R}_i}-{\bbox{R}_j})\, .
\end{equation}Please note that we use a different definition of $J$
that that used by Lovesey\cite{Lovesey_book}. The Fourier
transform of the spatial spin-spin correlations
$<\bbox{S}^{\alpha}_{i} \bbox{S}^{\alpha'}_{j}>$ is
\begin{equation}
<{\bbox{S}}^{\alpha}_{{\bbox{Q}}}
{\bbox{S}}^{\alpha}_{-{\bbox{Q}}}> \propto \frac{1}{3} S (S+1)
\left(1-\frac{1}{3}\beta S(S+1)J({\bbox{Q}})\right)\, .
\label{Eq_SQ}
\end{equation}

For a 1D magnet with a nearest-neighbor exchange $J$,
$J({\bbox{Q}})= 2 J \cos(\pi Q_c)$ and the second moment becomes
\begin{equation}\label{Eq_second_1D}
<\omega^2>=\frac{4}{3} \frac{S(S+1)}{1-\frac{1}{3}\beta S(S+1)
J(Q_c)} \cdot J^2(1-\cos(\pi Q_c))\nonumber
\end{equation}

\widetext
\begin{table}[t]
\caption{For neutron energy loss, the integrated intensity
$R_0(\bbox{Q})=\int_0^{\infty}d\omega S(\bbox{Q},\omega)$ and the
first and second moment $<\omega>(\bbox{Q})=\int_0^{\infty}d\omega
S(\bbox{Q},\omega)\omega/R_0(\bbox{Q})$ and
$<\omega^2>(\bbox{Q})=\int_0^{\infty}d\omega
S(\bbox{Q},\omega)\omega^2/R_0(\bbox{Q})$ of the observed spectra.
$R_0$ is in arbitrary units (a.u.) and $<\omega>$ and
$<\omega^{2}>$ are in units of meV and meV$^2$, respectively.
Columns 2 to 4 are derived from MARI data with a varying
wave-vector transfer perpendicular to the chain. The final columns
are derived from the RITA data with a different normalization.}
\vspace{0.2cm}
\begin{tabular}{|cc|ccc|ccc|c||cccc|}
&T(K)& & $Q_c=0.5$ & & & $Q_c=1$ & & $<{\mathcal H}>$ (a.u.)& &
$\bbox{Q}=(0.81,0.81,1)$ & &\\ \tableline

&&$R_0$ (a.u.)& $<\omega>$& $<\omega^{2}>$& $R_0$ (a.u.)&
$<\omega>$& $<\omega^{2}>$& & $R_0$ (a.u.) & $<\omega>$ &
$<\omega^{2}>$ &
\\\tableline

&6 & 6.1(0.9)& 5.8(0.3)& 6.0(0.3) & 22.2(0.7) & 2.6(0.1) &
3.3(0.1) & -28(0.4)& 1316(21) & 1.76(0.05) & 2.35(0.09)  &\\

&7 &  &  &  &  &  &  & & 1362(45) & 1.96(0.1) & 2.49(0.08) &\\

&8 & 5.2(1.1)& 5.4(0.5)& 5.6(0.4) & 20.7(0.9) & 2.57(0.17)  &
3.25(0.23) & -27(0.5)& 1323(44)& 2.06(0.08) & 2.57(0.05)  &\\

&9 &  &  &  &  &  &  &  & 1322(46) & 1.91(0.12) & 2.53(0.16) &\\

&10 &  &  &  &  &  &  &  & 1404(32) & 1.95(0.06) & 2.36(0.05) &\\

&12 & 4.3(1.0)& 5.8(0.6)& 5.9(0.4) & 20.84(0.66) & 2.89(0.6) &
3.58(0.22) & -26(0.5) & 1233(45)& 2.17(0.13) & 2.78(0.13)  &\\

&15 &  &  &  &  &  &  &  & 1055(42) & 2.17(0.16) & 2.52(0.23) &\\

&20 &  &  &  &  &  &  &  & 939(39) & 3.1(0.2) & 3.62(0.13) &\\

&25 & 6.1(1.0)& 5.0(0.4)& 5.4(0.3) & 16.55(0.48) & 3.72(0.2) &
4.37(0.24) & -21.3(0.2)& 813(37)& 3.04(0.22) & 3.66(0.15) & \\

&30 &  &  &  &  &  &  &  & 729(36) & 3.0(0.23) & 3.48(0.18) &\\

&35 &  &  &  &  &  &  &  &689(35) & 3.37(0.25) & 4.03(0.14) &\\

&50 & 5.3(0.9)& 4.4(0.3)& 4.9(0.3) & 10.89(0.26) & 4.03(0.12) &
4.49(0.1) & -14.2(0.2) & 641(29)& 3.75(0.23) & 4.43(0.09) &\\

&70 &  &  &  &  &  &  &  & 571(28) & 3.78(0.26) & 4.43(0.1) & \\

&200 & 5.2(0.9)& 3.5(0.2)& 4.0(0.2) & 7.5(0.19) & 4.46(0.14) &
4.95(0.12)  & -3.8(0.1) & &  & &
\\
\end{tabular}
\label{positive_moments}
\end{table}
\narrowtext
This result can be directly compared with the
measurements at $T=200\;\mathrm{K}$. The calculated $S(Q_c)$ given
by Eq.~\ref{Eq_SQ} with $J=2.28\;\mathrm{meV}$ is in excellent
agreement with the experiment at $T=200\;\mathrm{K}$
(Fig.~\ref{MARI-int-first-200K}) if an overall scaling factor is
fitted to the experimental data.\par

\begin{figure}
\begin{center}
  \includegraphics[height=9cm,bbllx=73,bblly=190,bburx=485,
  bbury=650,angle=0,clip=]{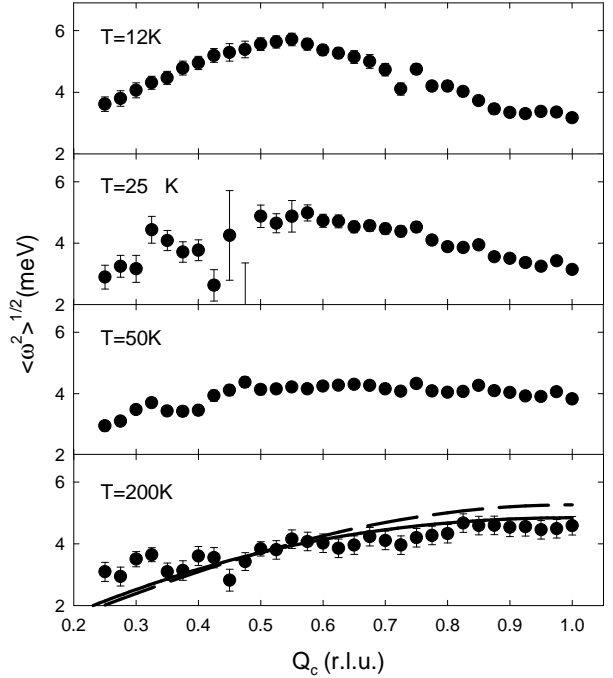}
  \vspace{0.1cm}
  \caption{Square root of the second energy moment $\Omega(Q_c)$
  for several temperatures as a function of wave-vector transfer
  $Q_c$. The solid line is the theoretical prediction given by
  Eq.~\protect\ref{Eq_second_1D}. The dashed line is the
  prediction for infinite temperature.}
  \label{MARI-2nd-moment}
\end{center}
\end{figure}
\newpage

A more sensitive test of the theory is a quantitative comparison
of the predicted and measured second energy moment. The square
root of the second moment $\Omega=<\omega^2>^{1/2}$ was determined
numerically from the measured spectra after the usual corrections
were made. Fig.~\ref{MARI-2nd-moment} shows that $\Omega(Q_c)$ has
a maximum which approaches $Q_c=1$ at high temperatures. The solid
line shows the predicted $\Omega(Q_c)$ according to
Eq.~\ref{Eq_second_1D} at $T=200\;\mathrm{K}$ and using
$J=2.28\;\mathrm{meV}$. There is good quantitative agreement
between theory and experiment considering that the solid line is
not a fit but a prediction based only on the temperature and the
exchange constant along the chain. $\Omega(Q_c)$ for infinite
temperatures $\beta=0$ is shown by the dashed curve for
comparison.\par

Fig.~\ref{MARI-2nd-moment} also shows that the maximum in
$\Omega(Q_c)$ changes from $Q_c=1$ to $Q_c=0.5$ on cooling. At
$T=12\;\mathrm{K}$, the maximum is where the dispersion has its
largest energy. The moments of the spin spectrum are summarized in
Table~\ref{positive_moments}. Here the integrals are done only
over the positive energy response.\par

\section{Conclusions}
In summary, the evolution of the dynamic spin structure factor
$S(Q_c,\omega)$ of ${\rm CsNiCl_{3}}$ has been surveyed between
$T=5$ and $200\;\mathrm{K}$ over a wide range of wave-vector
transfers $Q_c$ along the chain.\par

The temperature dependence of the correlation length $\xi(T)$ was
determined from the $Q_c$-dependence of the equal-time structure
factor and the results are in agreement with quantum Monte Carlo
calculations except for $T<10\;\mathrm{K}$, close to the ordering
temperature $T_N$. At these temperatures and at the non-critical
1D wave-vector $(0.81,0.81,1)$ the correlation length is shorter,
surprisingly, than is predicted for isolated chains, and the gap
is higher than expected. The intensities of the excitations do not
scale with the $1/\omega$-law which is typical for
antiferromagnets. These results are in sharp contrast with the
predictions of the random phase approximation and with the single
mode approximation.\par

The magnetic excitations remain resonant and dispersive up to
$T=50\;{\mathrm K} \simeq 2 J$, a temperature comparable with the
spin-band width. They remain resonant because the energy of the
spin excitations renormalizes upward more rapidly than their
damping. The temperature dependence of the dispersion along the
chain direction was measured, and the excitation energy
$\Delta(T)$ and the spin velocity $v_s(T)$ were determined for
$6\;\mathrm{K}<T<50\;\mathrm{K}$. We found that the relation
$\xi=\frac{\sqrt{v_{\rm s}^2+\alpha^2/4}}{\Delta}$, which is
predicted by the single mode approximation, remains valid in the
entire temperature region. This suggests that the excitation
spectrum around the antiferromagnetic point consists of
single-particle excitations even at $T=50\;\mathrm{K}$ where the
excitations are broad. It also hints at the existence of a hidden
spin order which carries the propagating excitations as predicted
by numerical calculations. The observed excitations may be
regarded as domain walls (solitons) of the hidden string order,
whose energy scale is $J$ rather than $\Delta$.\par

At $T=200\;\mathrm{K}$ the scattering is well described as
paramagnetic scattering and extends to $\sim 10\;\mathrm{meV}$
energy transfer at the antiferromagnetic point. The upper boundary
of the paramagnetic scattering exhibits a broad maximum at the
antiferromagnetic point.\par

The structure factor and the second energy moment are consistent
with the first-order high-temperature series expansion. These
results call for an extension to higher temperatures of the theory
for gapped weakly coupled spin chains.\par

\begin{acknowledgments}
We would like to thank Dr. O. Petrenko  for his assistance with
experiments at ISIS and Dr. P. Santini and Dr. D.~A. Tennant for
enlightening discussions. We are grateful to K.~N. Clausen and the
staff of Ris$\o$ National Laboratory for their expert assistance
and for the use of the excellent neutron facilities that made this
research possible, regrettably no longer available with the shut
down of DRB. Financial support for the experiments was provided by
the EPSRC, by the EU through its Large Installations Program and
by the British Council-National Research Council Canada Program.
ORNL is managed for the U.S. D.O.E. by UT-Battelle, LLC, under
contract no. DE-AC05-00OR22725. One of the authors (M.~K.) was
supported by the Swiss National Science Foundation under Contract
No. 83EU-053223.
\end{acknowledgments}

\bibliographystyle{prsty}

\end{document}